\documentclass[sigconf]{acmart}

\AtBeginDocument{%
  }

\copyrightyear{2022}
\acmYear{2022}
\setcopyright{acmcopyright}
\acmConference[SIGIR '22] {Proceedings of the 45th International ACM SIGIR Conference on Research and Development in Information Retrieval}{July 11--15, 2022}{Madrid, Spain.}
\acmBooktitle{Proceedings of the 45th International ACM SIGIR Conference on Research and Development in Information Retrieval (SIGIR '22), July 11--15, 2022, Madrid, Spain}
\acmPrice{15.00}
\acmISBN{978-1-4503-8732-3/22/07}
\acmDOI{10.1145/3477495.3532032} 

\usepackage{multirow}
\usepackage{dblfloatfix}
\usepackage{enumitem}
\usepackage{bbm}
\usepackage{balance} 

\newcommand{\IR}{$\mathcal{M}_{IR}$}

\begin{document}
\fancyhead{}

\title{Offline Evaluation of Ranked Lists using Parametric Estimation of Propensities}

\author{Vishwa Vinay}
\affiliation{%
    \institution{Adobe Research}
    \city{Bangalore}
    \country{India}
}
\email{vinay@adobe.com}

\author{Manoj Kilaru}
\authornote{Work done while at Adobe Research}
\affiliation{%
    \institution{University of California}
    \city{San Diego}
    \country{USA}
}
\email{mkilaru@ucsd.edu}

\author{David Arbour}
\affiliation{%
    \institution{Adobe Research}
    \city{San Jose}
    \country{USA}
}
\email{arbour@adobe.com}

\renewcommand{\shortauthors}{Vinay et al.}

\begin{abstract}
Search engines and recommendation systems attempt to continually improve the quality of the experience they afford to their users. Refining the ranker that produces the lists displayed in response to user requests is an important component of this process. A common practice is for the service providers to make changes (e.g. new ranking features, different ranking models) and A/B test them on a fraction of their users to establish the value of the change. An alternative approach estimates the effectiveness of the proposed changes offline, utilising previously collected clickthrough data on the old ranker to posit what the user behaviour on ranked lists produced by the new ranker \textit{would have been}. A majority of offline evaluation approaches invoke the well studied inverse propensity weighting to adjust for biases inherent in logged data. In this paper, we propose the use of parametric estimates for these propensities. Specifically, by leveraging well known learning-to-rank methods as subroutines, we show how accurate offline evaluation can be achieved when the new rankings to be evaluated differ from the logged ones.
\end{abstract}



\maketitle

\section{Introduction}
Many user-facing applications display a ranked list of items - this includes search results, ads, and recommendations. The design of such an application involves many choices, and the effect of each choice needs to be evaluated. In this paper, we restrict ourselves to factors affecting the ranked order of the displayed items - for example, features used as input into the ranker and learning-to-rank algorithm alternatives. Traditionally, the effects of these changes are evaluated using information retrieval metrics computed over ground truth relevance labels obtained from judges for a chosen set of queries. This \textit{Cranfield paradigm}, which has been used by tracks organized by TREC\footnote{Text REtrieval Conference: \url{https://trec.nist.gov/}}, allows the comparison of alternative rankers via evaluation on standardised test collections. There continues to be progress on this front - bridging the gap between system measures and utility for the user~\cite{lioma17, carterette2011} and more effective and efficient gathering of relevance labels~\cite{lu2017}. 

In parallel, online evaluation of ranking systems~\cite{onlineEvalIR} is an active area of research. A/B Testing~\cite{abTesting, kohavi2013} has been adopted as the de facto standard by practitioners wanting to verify the effect of changes to a ranker. Similar to offline approaches, current research focuses on more effective and robust evaluation metrics and methodologies~\cite{lihong2015, schuth2014}. Despite its successes, there are reasons why online A/B tests may not always be the best option. Primarily, there is the danger that a proposed change degrades the user experience, thereby negatively impacting the subset of users that were exposed to the change. 

Such concerns have led to recent research that looks at ``offline A/B tests''~\cite{ref_lncs10}. The central idea here is to utilise historical user interaction data - query logs, ranked lists of items displayed to the user, and clicked results. Making effective use of data collected by the engine allows us to mitigate the implementation costs of rolling out an A/B test and the risks of reduced user experience. Evaluation in this setting is posed as a counterfactual question - ``\textit{How would the user have interacted with the results if presented in the order proposed by the new ranker}?''. 

The reason this is not straightforward to answer is that an unbiased estimate of the effect of the proposed changes needs to be obtained using biased historical data. The collected data is biased because we only have user interactions on items presented at the previously decided rank positions. The new ranker might differ from the old one in terms of the set of results shown, as well as their ordering. Moreover, users are known to have inherent biases, like preferring results at higher ranks. The unbiased evaluation reflects the fact that we would like to obtain an estimate as close to the new ranker's true (unknown) performance as possible. 

This paper focuses on deterministic rankers -- models based on popular choices within the learning-to-rank setup (e.g., support vector machines, gradient boosting machines, and neural networks). Building on existing work that leverages a statistical technique known as \textit{Inverse Propensity Weighting} (IPW), we show how effective evaluation can be achieved via parametric estimates of the corresponding propensities. In our approach we train an \emph{Imitation Ranker} that mimics the observed rankings in historical clickthrough data. Given the trained imitation ranker, our approach provides a parametric estimation for the probability of a document being observed at a certain rank in response to a given query. This propensity can be used for the estimation of a relevance metric for a new ranking policy.

In our experimental setting, we show that existing IPW based evaluations consistently lead to lower values of the ranking metrics, i.e., the new ranker will be believed to produce rankings of lower relevance than their actual true value. Parametric propensities computed at the fine level of detail of document and rank pairs addresses this underestimation bias, but come at the cost of increased variance. We show results that aid in obtaining a robust (low variance) and accurate (low bias) evaluation. 

\section{Related Work}
\label{sec:RelatedWork}

The focus of the current paper is evaluation, but there are overlaps with the closely related area of \textit{Unbiased Offline Learning to Rank} which we enumerate in this section. Two early papers on this topic~\cite{wang2016, joachims2017} outlined the core problem that logged clickthrough data is incomplete. They suggest the use of a per-rank examination propensity to account for the position bias where results at higher ranks are more likely to be examined. Subsequent papers address the problem of estimating this rank-bias efficiently - by utilising randomisation experiments~\cite{wang2018position} or using natural variation observed in logged data~\cite{agarwal2018consistent, agarwal2019estimating, fang2019intervention, oosterhuis2021unifying}.

The topic continues to be a very active area of exploration - with papers looking at the optimisation of a wide class of metrics using larger capacity models~\cite{agarwal2019general, lambdaMart, ai2018unbiased}, as well as online versions of these algorithms~\cite{oosterhuis2018differentiable, oosterhuis2020unbiased}. Tutorials on the topic~\cite{wsdm2015Tutorial, sigir2016Tutorial, sigir2019Tutorial} summarise the current status and point to promising future directions.

The focus of the current paper coincides with recent work that considers biases beyond the position bias. We refer to the examination propensity as a \textit{user bias} to indicate the source of the incompleteness in the data. Another example of a user bias is the effect of differential trust for results at different ranks~\cite{agarwal2019addressing}. There is the related class of \textit{system biases} - where decisions made by the historical ranker need to be appropriately accounted for. An example is the issue of fairness of exposure~\cite{yadav2019fair} to counter a rich-get-richer phenomenon for items being chosen into impressions. Closest in motivation to the work described in the current paper is the notion of selection bias~\cite{ovaisi2020correcting}. Most unbiased learning-to-rank methods make the assumption that all documents are observed, and therefore debiasing only needs to account for the rank-specific effects. 

Our work considers the statistical proclivity of the historical ranker to place certain (types of) documents at certain ranks. We model this as a propensity computed at the level of a (document, rank) pair. Modelling biases at the (document, rank) level has the additional advantage that it allows us to target metrics where the user clicks on multiple items in the same impression. The next section details how the specifics of obtaining this propensity affects its downstream use within an IPW-based evaluation framework.

\section{Problem Statement and Methodology}

Consider a dataset collected by an operational search engine that follows policy $\pi$ for retrieving and ranking documents. This data is of the form $D_{\pi}={(q, I, c)}$ where:
\begin{itemize}[leftmargin=*,topsep=0pt]
\setlength\itemsep{-0.9em}
\item $q$ represents the user query \\
\item $I$ is the impression, an ordered list of $K$ documents returned in response to $q$ \\
\item $c$ is an array where element $c_{k}\in \{0, 1\}$ indicates if the document at rank $k$ was clicked
\end{itemize}
In this paper, we use the terms `Impression' and `List' to refer to a ranking over a small set of $K$ documents chosen from a large set of indexed items. $I_{k}=d$ provides the identifier of the document that was shown at rank $k$. 

The dataset $D_\pi$ may contain the same query string multiple times, and the observed impressions across the occurrences of $q$ may have variations. This is natural when the ranker $\pi$ is stochastic~\cite{diaz2020evaluating}, but some degree of variation is natural even for deterministic rankers. For example, rankers typically contain feedback loops with clickthrough-based features or personalized and context-aware features which can potentially alter the rankings in response to user behaviour. We refer to the set of impressions seen in response to $q$ as $\pi(q)$. 

The central task considered in this paper is evaluating the results generated by a new ranker $\mu$. We represent the results generated by $\mu$ in response to a query $q$ by $\bar{I} = \mu(q)$. Each $\bar{I}$ is evaluated with respect to a relevance metric $M(\bar{I}, c)$ for a given impression and available click labels. In this paper, we focus on impression-level relevance metrics that can be decomposed in an additive manner over individual items: $M(I, c) = \sum_{k=1}^K m(c_k, k)$

This restriction translates to having the relevance of an impression be an aggregation of click events observed on individual items. Note that the constraint is not very limiting, and includes a wide variety of IR metrics~\cite{ref_lncs1}. In this paper, we consider the following two:
\begin{enumerate}
    \item Number of Clicks: $NoC(I,c) = \sum_{k=1}^K c_k$
    \item Mean Reciprocal Rank: $MRR(I,c) = \frac{1}{K} \sum_{k=1}^K \frac{c_k}{k}$
\end{enumerate}

The metrics might differ in the ``gain'' (directly taken to be $c_k$ in the above equations) or the rank-specific ``discount'' ($MRR$ uses $k$, but $NDCG$ would take $log(k)$). The methods presented in this work can be readily applied to the estimation of a wider class of ranking metrics of interest - including Precision@K, MAP and NDCG. Note that the definition of MRR above is a non-standard generalisation to handle multiple relevant (clicked) documents for a given query. The \textit{value} of a ranker $\mu$ -- given by $V(\mu)$ - represents the expected value of the relevance metric for impressions produced by $\mu$. The task is to provide an accurate estimation of $V(\mu)$ by evaluating rankings that $\mu$ produces for each query $q \in Q$.

\subsection{Background}
When $\mu$ produces results different from $\pi$, we might end up with items for which we do not have click information. That is, it might be the case that $\bar{I} \neq I$. One option is to restrict ourselves to the $q$ where $\bar{I}=I$, i.e., when the old and the new ranker produce the same rankings. This gives us one estimate of $V(\mu)$:
\begin{equation}
\label{eq:One}
\hat{V}(\mu) = \frac{1}{|D_{\pi}|} \sum_{(q,I,c) \in D_{\pi}} \mathbbm{1}\left\{\bar{I}=I\right\}M(I,c)
\end{equation}

$\hat{V}(\mu)$ provides a biased view of the true $V(\mu)$ because of the logging policy confirming behaviour that it exhibits. That is to say, any $\mu$ that consistently produces rankings that are very similar to those in the historical impressions produced by $\pi$ is likely to have a relatively higher estimate $\hat{V}(\mu)$. Conversely, if the new ranker's results vary from those in the logged data, the estimated value $\hat{V}(\mu)$ from Equation~\ref{eq:One} will be an underestimate. Addressing this bias is at the core of the problem. 

One way to conduct an accurate evaluation of $\mu$ over the logged dataset $D_{\pi}$ is via the use of the well-studied Inverse Propensity Weighting (IPW) mechanism~\cite{ref_lncs5}. An IPW version of Equation~\ref{eq:One}, where the relevance metric term has been replaced by its summation equivalent is:
\begin{equation} 
\label{eq:ListIPW}
\hat{V}_{L}(\mu) = \frac{1}{|D_{\pi}|} \sum_{(q,I,c) \in D_{\pi}}  \frac{\mathbbm{1}\left\{\bar{I}=I\right\}}{\hat{p}(\bar{I}|q)}\sum_{k=1}^K m(c_k,k)
\end{equation}

Propensity here refers to the term $\hat{p}(I|q)$, which represents how likely it is that the old ranker $\pi$ returned impression $I$ in response to query $q$. The purpose of this propensity is to re-weight the examples so as to simulate the situation where $D_{\pi}$ was collected in an experimental setting. 

The subscript $L$ in $\hat{V}_{L}$ indicates that the propensity is computed at a list level and is equivalent to the method described in~\cite{replay}. 
One downside of the list estimator, is that it is statistically inefficient - a significant fraction of $D_{\pi}$ will be discarded if we are looking for exact matches at the list level (i.e., the indicators in Equations~\ref{eq:One}~\&~\ref{eq:ListIPW}). In order to make better use of the logged data, methods that utilise document-level matches have been designed. In particular,~\cite{li2018offline} suggests the factorisation of an impression using click models~\cite{chuklin2015click}. In the current paper, we utilise the ``Item+Position Click Model'' proposed by~\cite{li2018offline}, which suggests that observations for a (document, rank) combination -- $O(d,k|q)$ -- requires the pair to be modelled jointly. Well known click models can be derived from the above general form by making additional assumptions. For example, the Position-Based Model~\cite{craswell2008experimental} is obtained by setting $O(d,k|q) = R(d|q) * E(k)$, a relevance-only component and a per-rank examination factor. 

Under the Item+Position Click model, an IPW estimator can be obtained as:
\begin{align}
\label{eq:PosItemIPW}
\hat{V}_{IP}(\mu) = \frac{1}{|D_{\pi}|} \sum_{(q,I,c) \in D_{\pi}} \sum_{k=1}^K  \frac{\mathbbm{1}\left(\bar{I}_k=I_k\right)}{\hat{p}(\bar{I}_k, k | q)} m(c_k,k)
\end{align}

Unlike the List estimator ($\hat{V}_{L}$), we note that the matching operates at the level of individual documents. Therefore, when $\mu$ places a document $d$ at rank $k$ for query $q$, any historical impression that contains this $\{d,k,q\}$ tuple contributes to the estimated value (even if there are differences in documents at other ranks). The corresponding propensity, $\hat{p}(\bar{I}_k,k|q)$, helps re-weight the historical data to account for the non-uniform likelihood of $\pi$ placing certain documents at particular ranks. This propensity is the main object of interest in the current paper. Since the term $\hat{p}(\bar{I}_k,k|q)$ includes the document ($\bar{I}_k=d$) and the rank ($k$), we refer to it as a (document, rank) propensity.

Existing work mostly only considers a rank-specific propensity $\hat{p}(k)$. That is, compared to the model described above, this is equivalent to making the assumption that a given document is equally likely to be present at all ranks in the logged historical dataset. While this assumption is unlikely to hold in general, it leads to an advantage that the number of propensities to be estimated is only $\mathcal{O}(K)$. For the proposal considered here, the full set of (document,rank) propensities to be estimated is $\mathcal{O}(N_QK^2)$, where $N_Q$ is the number of queries we wish to evaluate over. Our challenge is to therefore have a robust mechanism to estimate these propensities so as to benefit from the increased model richness.

To do this, we assume access to a set of features for a query-document pair, we denote this representation by $x_{qd}$. Note that the rankers $\pi$ \& $\mu$ would typically utilise a set of features to produce per-document relevance scores, with impressions formed by ordering over these items and their scores. We do not require that the set of features $x_{qd}$ be the same as those used by either $\pi$ or $\mu$. There is recent precedence for the use of features within an evaluation setting (e.g.~\cite{wang2018position}).
Our primary concern is to define a methodology that allows us to rely on experimental data gathering to a minimal extent. We expect the features to provide generalisation when there are systematic differences between the ranker that produced the logged dataset and the new ranker that is to be evaluated.

\begin{figure*}[!ht]
\centering
\includegraphics[width=\textwidth]{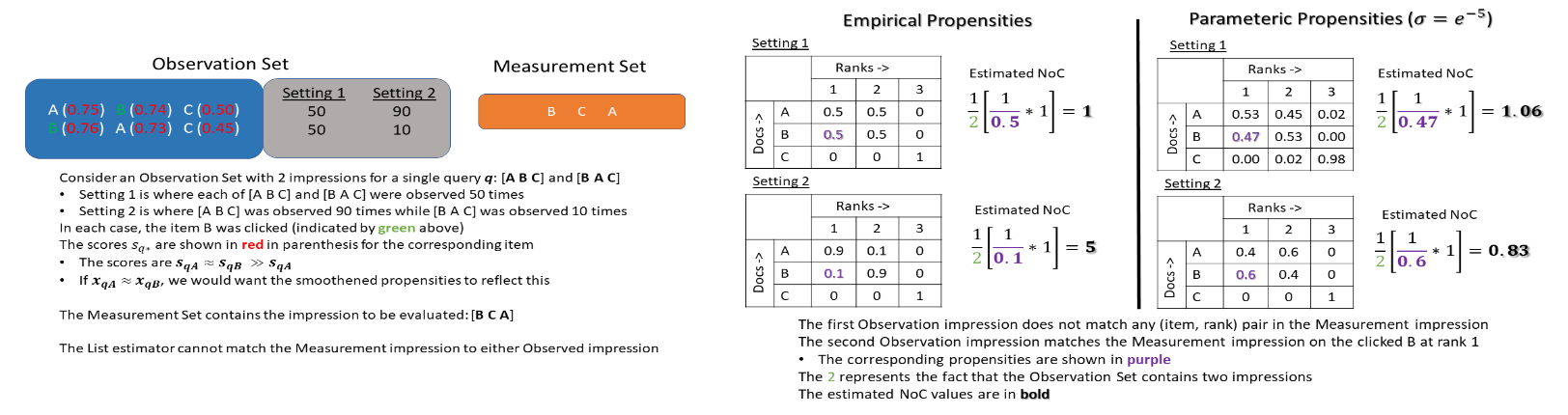}
\caption{A toy example illustrating the approach described in this paper. The logged Observation Set $D_{\pi}$ contains a single query $q$ for which $\pi$ had returned two different orderings at different times -- $[A B C]$ and $[B A C]$. Click information is available for the observed data -- document $B$ was always clicked. We have been asked to estimate the number-of-clicks (NoC) that might be observed on an impression produced by $\mu$-- $[B C A]$ -- for the same query . When the new ranker contains a (document,rank) pair that is rare in the historical data, the existing methods though unbiased produce unreasonably large estimates of the metric. E.g. ``Setting 2'' where $B$ was returned at rank $1$ only $10\%$ of the time in $D_{\pi}$ leads to an estimated NoC=$5$. Using parametric estimates of the propensities takes advantage of similarities of items in feature space. This in turn leads to a more balanced evaluation for (document,rank) combinations that are rare in the observed data but explained by the features}
\label{fig:exampleFigure}
\end{figure*}

\subsection{Computing (Document, Rank) Propensities}
One option is to use the empirical propensities:
\begin{equation}
\label{eq:EmpProps}
\hat{p}(d, k | q) = \frac{\sum_{(q',I,c) \in D_{\pi}} \mathbbm{1}\left(I_k=d\right)\mathbbm{1}\left\{q'=q\right\}}{\sum_{(q',I,c) \in D_{\pi}} \mathbbm{1}\left\{q'=q\right\}}
\end{equation}

This equation estimates the propensities as the fractional number of times a particular document $d$ was shown at rank $k$ across all impressions for the query $q$ in $D_{\pi}$. Despite the presence of some variation in rankings for the same query, the set of propensities obtained from Equation~\ref{eq:EmpProps} is likely to be very sparse. A $\{d,k,q\}$ tuple that is rare in $D_{\pi}$ will lead to a very large inverse propensity weight when used in Equation~\ref{eq:PosItemIPW}. If the ranking that $\mu$ produces matches this rare tuple, the estimated $\hat{V}_{IP}(\mu)$ will have a large value.

A core contribution of this paper relies on leveraging standard learning-to-rank (LtR) methods to obtain the required propensities. Our proposed workflow has two components:
\begin{itemize}[leftmargin=*]
    \item \textbf{Training an Imitation Ranker:} Given a logged set of impressions ($D_\pi$) and a set of features for query-document pairs ($x_{qd})$, we train a model that produces scores ($s_{qd}=f(x_{qd})$) that allows the re-creation of the impressions produced by $\pi$ as accurately as possible. 
    \item \textbf{Computing Rank Distributions:} Given the set of $K$ scores for an impression from the trained Imitation Ranker, we fill out a $K \times K$ matrix where the entry at $(d,k)$ provides the propensity for item $d$ being placed at rank $k$.
\end{itemize}
We have assumed that the historical dataset $D_{\pi}$ only contains information about document rankings and clicks. If the score produced by $\pi$ for a query-document pair were logged as part of $D_{\pi}$, we could use them directly in the second step for computing the (document,rank) propensities. Because they are derived from a parametric model, these propensities are expected to be smoother than the ones in Equation~\ref{eq:EmpProps}. Figure~\ref{fig:exampleFigure} provides an illustrative example. 

\subsubsection{\textbf{Training an Imitation Ranker}}
\label{seq:IR}
The Imitation Ranker (\IR) is a function $f$ that produces a score $s_{qd}=f(x_{qd})$ given the features of the query-document pair as input. The features can include hand-crafted features or latent features from deep learning models. We propose the minimisation of the RankNet~\cite{burges2005learning} objective to train~\IR:
\begin{align}
\label{eq:pairwiseObj}
\mathcal{L}_{\text{pairwise}} = \sum_{q \in Q} \sum_{I \in \pi(q)} \left[\sum_{(d,z) \in I : d \triangleright z} \log(1+\exp{(s_{qd}-s_{qz})})\right]
\end{align}
where the set $Q$ contains all the queries in the logged dataset. The document pairs $(d,z)$ are chosen from the original impression $I$ such that $d$ was ranked higher than $z$ by the logging policy $\pi$. The objective above encourages~\IR~to produce scores that respect the original observed orderings. Since the observed rankings are used as labels, this step is similar to training a ranking model via knowledge distillation~\cite{hinton2015distilling} or weak supervision~\cite{dehghani2017neural}. In our experimental section, we evaluate what effect the choice of objective function has towards the end goal of producing a reliable evaluation.

\subsubsection{\textbf{Computing Rank Distributions}}
\label{seq:RankDistributions}
 
Let $s_{qd}$ be the score of the document $d$ for the query $q$ produced by the imitation ranker \IR. A historical impression with $K$ items leads to an array of $K$ scores.
We make a Gaussianity assumption for the scores: $p(s_{qd}) = \mathcal{N}(S_{qd} ; s_{qd}, \sigma^2)$, and define a pairwise contest probability $p_{dz}$ of document $d$ being ranked higher than $z$. The log-likelihood of the observed data $D_{\pi}$ over rankings produced by~\IR~can be defined as:
\begin{align}
\mathcal{L}_{\sigma} &= \sum_{q \in Q} \sum_{I \in \pi(q)} \left[\sum_{(d,z) \in I : d \triangleright z}\log(p_{dz})\right] \nonumber\\
 &= \sum_{q \in Q} \sum_{I \in \pi(q)} \left[\sum_{(d,z) \in I : d \triangleright z} \log\left(p(s_{qd}-s_{qz} > 0)\right)\right] \nonumber\\
 &= \sum_{q \in Q} \sum_{I \in \pi(q)} \left[\sum_{(d,z) \in I : d \triangleright z} \log\left(\int_0^{\infty} \! \mathcal{N}(s ; s_{qd}-s_{qz}, 2\sigma^2)\mathrm{d}s\right)\right]
 \label{eq:softRank2}
\end{align}

where the quantity $\sigma$ represents our uncertainty in the value of the score. This formulation was defined in SoftRank~\cite{taylor2008softrank}, which also provides the backbone of our proposed method. 

If we see $\mathcal{L}_{\sigma}$ as a function of only $\sigma$ (the document scores are given by \IR), we can infer the value of $\sigma$ that maximises the log-likelihood in Equation~\ref{eq:softRank2}. Intuitively, we can expect that if scores produced by \IR~reflect the logged impressions accurately, a smaller value of $\sigma$ will suffice. On the other hand, if the scores disagree with the logged rankings, a larger value of $\sigma$ will be inferred - this in turn will lead to $p_{dz}$'s that are closer to $0.5$ (i.e., uncertainty in pairwise orderings). Note the similarity between $\mathcal{L}_{\sigma}$ and the likelihood in Equation~\ref{eq:pairwiseObj} -- while $\mathcal{L}_{\text{pairwise}}$ uses the sigmoid of the score difference for the pairwise comparison, $\mathcal{L}_{\sigma}$ follows from SoftRank's use of Gaussians for the scores.

Let $W \in \mathbb{R}^{K \times K}$ be a $K \times K$ matrix, where both the rows and the columns are associated with documents and each element $W_{dz}$ is set to $p_{dz}$. From this matrix, we utilise the following recursion mechanism to derive a matrix $\bar{W} \in \mathbb{R}^{K \times K}$ where the rows are still associated with documents while the columns are associated with ranks. For initialisation, we set $\bar{W}^{(1)}_{dk}=\delta(k): \forall 1 \leq d \leq K$ where $\delta(x)=1$ when $x=1$ and zero otherwise. For $1 \leq k \leq K$ and $2 \leq t \leq K$, we recursively update the matrix by considering each $z \neq d$ in turn as follows:
\begin{equation}
\label{eq:softRank3}
\bar{W}^{(t)}_{dk} = p_{dz}\bar{W}^{(t-1)}_{d,k-1} + (1-p_{dz})\bar{W}^{(t-1)}_{dk}
\end{equation}
In the above equation, the item $d$ is referred to as the anchor with whom all other documents $z$ are compared. Once the recursion is complete, we normalise the rows and columns such that the $K*K$ matrix is now doubly stochastic - the entries for a given row (associated with documents) provide a distribution over ranks, while traversing a column (rank) is a distribution over documents. We utilise the resulting value of $\bar{W}^{(K)}_{dk}$ as the propensity $\hat{p}(I_{k}, k | q)$ in Equation~\ref{eq:PosItemIPW}, where $d$ is the item indexed by $I_k$.

As with the Gaussian assumption for the scores, the recursive method described in Equation~\ref{eq:softRank3} has been borrowed from the SoftRank model~\cite{taylor2008softrank}. We direct the readers to the work of Taylor \textit{et al} for more detail about its conventional use as a method to produce differentiable approximations to IR metrics. Our innovation in this paper is to utilise the same mechanism to produce smooth (document,rank) propensities for counterfactual offline evaluation. The use of~\IR~leads to two documents with similar features having similar (document,rank) propensities for the same query. As indicated in Figure~\ref{fig:exampleFigure}, we expect this to have a beneficial effect when the new ranker to be evaluated ($\mu$) produces an ordering that was rare in the logged dataset $D_\pi$ but plausible given the features $x_{qd}$.

\medskip

\noindent \textbf{Example:} To better illustrate our (document,rank) propensity computation approach, we provide a toy example with a query $q$ and three documents $A$, $B$, and $C$. Consider the second observation impression in Figure~\ref{fig:exampleFigure} $[B A C]$, let the scores $s_{qB}=0.76$, $s_{qA}=0.73$~\&~$s_{qC}=0.45$. For computing the rank distribution for $d=B$ as the anchor, we start with an array $\bar{W}^{(1)}(B,\cdot) = [1, 0, 0]$, where each of the three entries provides the current belief that $B$ should be placed at the corresponding rank. In the first iteration of the recursion, $t=2$, the rank distribution for $d=B$ is updated by comparing it with $z=A$. For $B$ to stay at rank $1$, we have $\bar{W}^{(2)}(B,1)=1*p_{BA}$, i.e., the probability of being at rank $1$ from the earlier iteration times the probability that document $B$ \textit{beats} document $A$. The pairwise contest probability is given by $p_{BA} =  \int_0^{\infty} \! \mathcal{N}(s ; s_{qB}-s_{qA}, 2\sigma^2)\mathrm{d}s$ as defined earlier. Using $\sigma=e^{-5}$ for illustration purposes leads to $p_{BA} = 0.602$. With probability $(1-p_{BA}) = p_{AB}$, document $B$ loses the pairwise contest and drops to rank $2$. This yields the updated rank distribution $\bar{W}^{(2)}(B,\cdot) = [0.602, 0.398, 0]$. In the next iteration of the recursion, document $B$ is compared against $z=C$. Since the scores $s_{qB} \approx s_{qA} \gg s_{qC}$, given the low uncertainty $\sigma$ in the scores, the pairwise contest probabilities $p_{AC} \approx p_{BC} \approx 1$. Thus, the final rank distribution of document $B$ remains similar to above, i.e., $[0.602, 0.398, 0]$. Therefore, the propensity of $B$ being at rank $1$ is $\hat{p}(B, 1 | q) \simeq 0.6$,  as shown on the right of Figure~\ref{fig:exampleFigure}. 

\subsubsection{\textbf{Summary}}
As a recap, given a logged dataset $D_{\pi}$, we follow the following four steps:
\begin{enumerate}
    \item We train an Imitation Ranker \IR~that best approximates the logged rankings as described in Equation~\ref{eq:pairwiseObj}.
    \item Given the scores produced by the trained \IR, we find the optimal $\sigma$ that maximises the log-likelihood in Equation~\ref{eq:softRank2}.
    \item Combining the scores $s_{qd}$ and the inferred $\sigma$, we compute rank distributions per document using the recursive algorithm described in Equation~\ref{eq:softRank3}.
    \item The rank distributions are the (document, rank) propensities to be used in an IPW setup as in Equation~\ref{eq:PosItemIPW}.
\end{enumerate}

Note that steps $(3)~\&~(4)$ are invoked for every impression produced by the ranker $\mu$ being evaluated. The training of \IR~and the inferring of an optimal $\sigma$ is a one-time operation for a given $D_{\pi}$.

In the next section, we conduct a series of experiments to illustrate the behaviour of our proposed method on the unbiased offline evaluation task. As previously mentioned, our primary contribution in this paper is the proposal to use parametric estimates of (document, rank) propensities for offline evaluation. We have described one possible workflow to derive these propensities, other alternatives may be possible, we leave the exploration of these to future work. 

\section{Experiments}
\label{sec:Experiments}

In the preceding sections, we have described a method that computes propensities for a (document,rank) pair given a query. In the current section, we setup a simulation study of different aspects of our workflow - highlighting where the proposal leads to benefits and where further work is required.

\subsection{Research Questions}
\label{sec:RQ}

We first enumerate the main axes of our investigation, and design experiments specifically to answer each of the following questions.

\noindent \begin{enumerate}
    \item[\textbf{RQ1}] When do existing methods fall short?~:~We expect the method described here to be particularly helpful in situations where there are systematic differences between the ranker that logged the dataset and the one being evaluated.
    \item[\textbf{RQ2}] How does the training of the~\IR~affect the results of the evaluation?~:~The~\IR~is a machine learnt model trained to reproduce observed rankings. We would like to correlate the generalisation ability of~\IR~with its role in an evaluation pipeline.
    \item[\textbf{RQ3}] What is the effect of observed rankings inconsistent with the features available to~\IR?~:~The ranker $\pi$ that led to the logged dataset is potentially a complicated function with a large set of features. Not all of these may be available to~\IR, and this would limit its ability to reproduce the rankings.  
\end{enumerate}

\subsection{Data and Experimental Setup}
\label{sec:Dataset}

\begin{figure*}[!ht]
\centering
\includegraphics[width=\textwidth,height=0.19\textwidth]{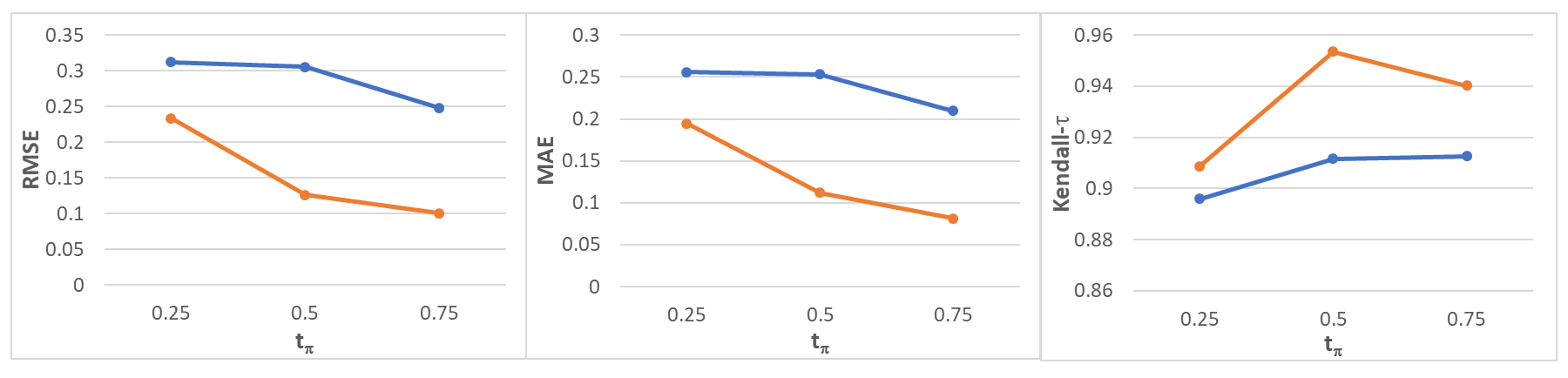}
\caption{Comparing the Measurement Ranker ($\mu$) with the Observation Ranker ($\pi$). In each case, the blue and orange lines represent $C=0.1$ and $C=0.001$ respectively}
\label{fig:obsMsmtCompare}
\end{figure*}

All the experiments described in the current paper are based on the Yahoo Learning to Rank dataset, specifically the larger \textit{Set $1$}, a publicly available collection popularly used in the LtR context. Since it does not come with clickthrough information, we generate user behaviour signals, the procedure for which is described in a subsequent paragraph. The dataset comes pre-divided into Train, Validate and Test splits which we utilise as follows.

We have two rankers - $\pi$ (which we refer to as the ``Observation Ranker'') and $\mu$ (referred to as the ``Measurement Ranker'') both trained using different samples from the Train split. We denote the fraction of the data that the training of each of the rankers is based on by $t_\pi$ and $t_\mu$. Since $\pi$ and $\mu$ are trained on different data, we can expect some level of differences between them. 

We convert the explicit relevance labels that the Yahoo dataset comes with into binary labels - ratings that are $3$ or above are replaced with a $1$, and a $0$ label otherwise. Both our rankers are $SVM^{rank}$ models~\footnote{\url{http://www.cs.cornell.edu/people/tj/svm_light/svm_rank.html}} trained using the default set of parameters except for the regularisation setting $C$. Emphasising the need for reducing the training error by setting $C$ to a low number results in models that are similar irrespective of the specific data used to train them - this is the effect of over-regularisation. We therefore utilise the $C$ parameter as a mechanism to control for differences between the two rankers. 

Once $\pi$ and $\mu$ are trained, we use the Validation split of the Yahoo dataset to quantify differences between the two rankers. We compute the Root Mean Squared Error (RMSE) and Mean Absolute Error (MAE) as score-centric differences between $\pi$ \& $\mu$. And, Kendall-$\tau$ provides signal on the differences between the orderings produced by the rankers. We also use the validation split for model selection, with the validation metric being the Kendall-$\tau$ between the model scores and simulated click labels (described below). While we have used RMSE, MAE and Kendall-$\tau$ in this paper, we could incorporate a wider set of metrics used by studies of IR system reproducibility~\cite{breuer2020measure} to quantify the differences between the two rankers $\mu$ and $\pi$, and study how these affect the robustness and reliability of the evaluation.

We produce top-$K$ lists from the two rankers for each query in the Yahoo Test split. We refer to the set of impressions produced by $\mu$ on this test split as the ``Measurement Set'' -- we would like to evaluate the Measurement Ranker based on the $\sim 3400$ impressions in this set. We take the set of unique queries in the Test split and construct a bootstrapped sampled-with-replacement set where each query is over-sampled proportional to the number of explicitly labeled relevant labels available for that query. The ranked lists produced by $\pi$ on this expanded set of queries leads to an ``Observation Set'' with $\sim 50,000$ impressions - together with simulated clicks, this represents the logged dataset $D_{\pi}$.

In Figure~\ref{fig:obsMsmtCompare}, we plot each of the $3$ metrics comparing our rankers ($\pi$ and $\mu$) as a function of two factors - ($1$) the regularisation parameter $C=\{0.1, 0.001\}$; and ($2$) the observation ranker training fraction $t_\pi=\{0.25, 0.50, 0.75\}$ on the X-axis. For these plots, the Measurement Ranker training fraction was set to $t_\mu=0.5$. The numbers in these plots, as well as in all subsequent experiments, are an average over $5$ runs.

With increased regularisation (lower values of $C$, the orange lines), we observe that the two rankers become more similar. This trend is present for both the score metrics (where lower values indicate higher similarity), as well as the ranking metric (Kendall-$\tau$). For a given value of $C$, we see that a larger value for $t_\pi$ leads to the rankers' behaviours being more similar. As a reminder, we wish to have a mechanism that controls the degree of similarity between the impressions generated by the two rankers to investigate the effect of the mismatch on the evaluation process. 

Following the method described in~\cite{joachims2017}, we simulate clicks on the ranked lists in the Observation Set as follows. The parameters controlling click noise are set to $\epsilon_{+}=1.0$ and $\epsilon_{-}=0.1$ -- a relevant item is always clicked while an irrelevant item receives a click with probability $0.1$. We set the position bias parameter $\eta=0.0$ to engineer a situation where the simulated clicks are not a function of rank. We simulated clicks on the Measurement Set impressions that we want to evaluate - metrics computed on these clicks provide the ground-truth $V_{\mu}$. We hold the clicks back and produce estimates of $\hat{V}_{\mu}$ using Equations~\ref{eq:ListIPW} and~\ref{eq:PosItemIPW}. An accurate evaluation requires that the estimate $\hat{V}_{\mu}$ be close to $V_{\mu}$.

Setting $\eta=0.0$ in our experimental setup is equivalent to an assumption that users examine all ranks equally - i.e., the rank-specific inverse propensity weights in~\cite{wang2016, joachims2017} will effectively be $1$. We show that in this setting, mismatches between rankings produced by $\pi$ and $\mu$ will produce an incorrect evaluation. A more comprehensive evaluation protocol will require the user and system biases to be jointly accounted for (e.g.~\cite{ovaisi2021propensity}). By controlling one of these factors (and deferring to existing work when user bias needs to be modelled), we ensure that the system biases - the focus of the current paper - are the main factors governing the experimental results. Section~\ref{sec:ParameterSensitivity} verifies the superior performance of our method with alternate clicks simulation parameter settings.

\subsection{Results and Analysis}

In this section, we consider the research questions outlined in~\ref{sec:RQ}, and design a series of experiments to answer them.

\subsubsection{Empirical Vs Parametric Propensities}
\label{sec:ListVsEmpProps}

\begin{table}[h!]
  \begin{center}
    \begin{tabular}{|l|l|l|c|c|c|c|} 
    \hline
    Metric & C & $t_{\pi}$ & GT & List & EP & IR \\ 
    \hline
    \multirow{6}{*}{NoC} & 
    \multirow{3}{*}{0.1} & 
         0.25 & 3.781 & 0.042 & 1.982 & 2.514 \\
     & & 0.50 & 3.772 & 0.146 & 2.204 & 6.202 \\
     & & 0.75 & 3.785 & 0.123 & 2.198 & 4.989 \\
    \cline{2-7} 
    & \multirow{3}{*}{0.001} & 
         0.25 & 3.586 & 0.107 & 1.988 & 2.904 \\
     & & 0.50 & 3.614 & 0.298 & 2.366 & 3.758 \\
     & & 0.75 & 3.595 & 0.277 & 2.391 & 3.331 \\
    \cline{1-7} 
    \multirow{6}{*}{MRR} & 
    \multirow{3}{*}{0.1} & 
         0.25 & 1.524 & 0.020 & 1.078 & 1.253 \\
     & & 0.50 & 1.519 & 0.063 & 1.146 & 2.036 \\
     & & 0.75 & 1.516 & 0.054 & 1.143 & 1.690 \\
    \cline{2-7} 
     & \multirow{3}{*}{0.001} & 
         0.25 & 1.450 & 0.048 & 1.065 & 1.345 \\
     & & 0.50 & 1.458 & 0.129 & 1.166 & 1.582 \\
     & & 0.75 & 1.451 & 0.117 & 1.187 & 1.504 \\
     \bottomrule
    \end{tabular}
    \caption{Comparison of the IPW-based estimators with the Ground Truth (GT): List-level matching leads to very low estimates, while the three Item-level alternatives are closer to the GT. EP=`Empirical Propensities'; IR=`Imitation Ranker'}
    \label{tbl:BasicResults}
  \end{center}
\end{table}

Our first experiment is designed to illustrate the benefits of deriving propensities from smoothened rank distributions, as well as highlighting the shortcomings of list-level matching. I.e., we wish to answer \textbf{RQ1}. Due to reliance of the IPW process on \textit{matching} a measurement data point with an equivalent one in the observation set, we would prefer methods that are less brittle in the presence of differences. 

We consider the measurement ranker trained using $50\%$ of the Train dataset, i.e., $t_{\mu}=0.5$, and we vary $t_{\pi} \in \{0.25, 0.5, 0.75\}$. As in Figure~\ref{fig:obsMsmtCompare}, we have two settings for $C \in \{0.1, 0.001\}$. We compute the Ground Truth (referred to as `GT' in the tables with experimental results) for each of the two metrics -- Number of Clicks (NoC) and Mean Reciprocal Rank (MRR). 
The first observation to be made from Table~\ref{tbl:BasicResults} is that even with minimal differences between rankings provided by the two rankers, List level matching provides incorrect estimates significantly lower than the GT. That is, despite what the high values of Kendall-$\tau$ in Figure~\ref{fig:obsMsmtCompare} indicate, the presence of the exact same impression in the Observation and Measurement sets is rare. Due to its much lower performance, we omit the results for the List estimator from future tables. Item-level matching displays much more stability, and the column titled `EP' -- standing for `Empirical Propensities' -- represents the method proposed in~\cite{li2018offline}. This is the combination that uses the propensities from Equation~\ref{eq:EmpProps} in the estimator defined in Equation~\ref{eq:PosItemIPW}. 

For~\IR, we use a $2$-layer neural net with tanh as the nonlinearity, and the model was trained for $500$ epochs. The input to~\IR~is the $700$-dimensional feature vector available with the Yahoo dataset. The propensities obtained by following the method described in Sections~\ref{seq:IR}~\&~\ref{seq:RankDistributions} lead to estimated values for NoC and MRR in the column titled `IR'. Smaller values of $C$ lead to the rankings produced by $\pi$ and $\mu$ to be similar, leading to EP being a competitive option. It is when the ranker to be evaluated produces impressions that are different to the one that produced the logged dataset that the benefits of parametric propensities becomes clearer. Note that countering the logging policy confirming behaviour was an explicit objective of ours, and therefore the use of features and an imitation ranker based setup for unbiased evaluation is our main contribution in this paper.

We note that for some combinations (e.g. $C=0.1$ and $t_{\pi}=0.5$), the estimated value of the metric in the IR column is larger than the GT. This behaviour is due to  small magnitude of the propensities - since it appears in the denominator of the estimators, this manifests as an unusually high value of the corresponding reward. The result is a high variance estimate of the value of $\mu$ - we are therefore trading-off an increased variance against the high bias estimate provided by the List estimator. There are well known prescriptions to handle large propensity weights, which our subsequent experiments employ.

\begin{figure*}[!h]
\centering
\includegraphics[width=.7\textwidth,height=0.2\textwidth]{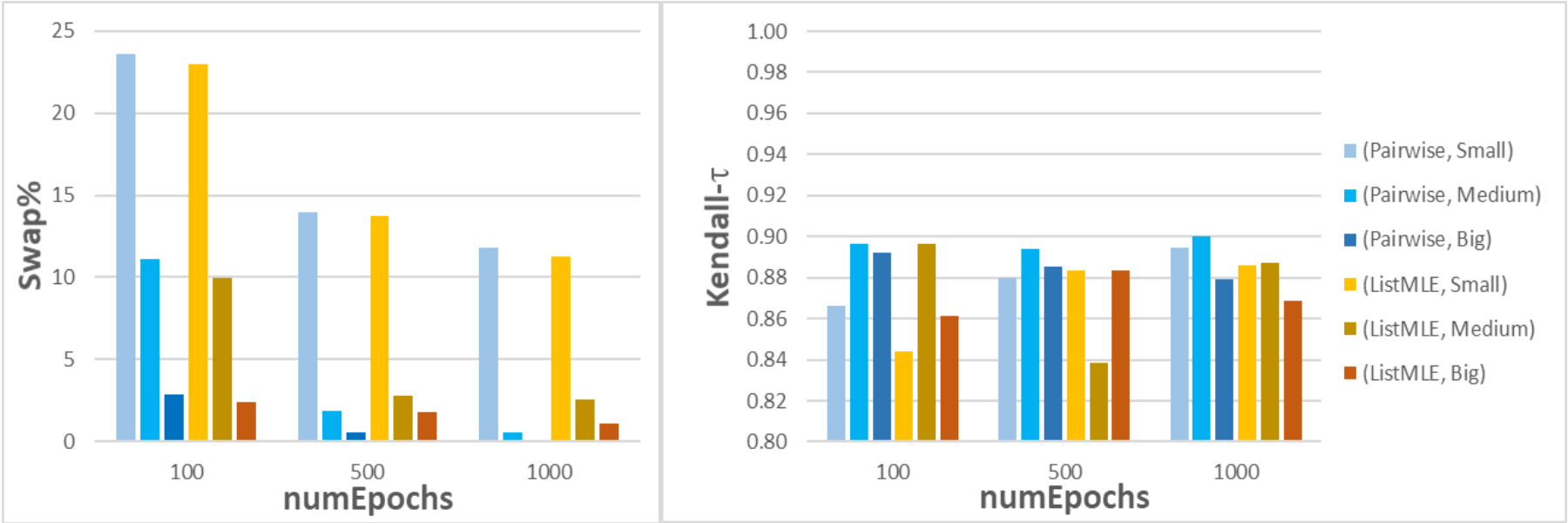}
\caption{Ranker metrics for different model architectures and ranking objectives}
\label{fig:modelObjNumEpochs}
\end{figure*}

\subsubsection{Training of the Imitation Ranker}
\label{sec:IRModel}

In the previous section, we trained an Imitation Ranker (\IR) to produce scores that reproduce the observed rankings. Rank distributions derived from these scores then fed into an IPW-based evaluation setup. We now look at the choice of objective used to train \IR, as well as the architecture of the model. That is, we address research question \textbf{RQ2}. 

We would like to setup a process where the evaluation methodology proposed here can take advantage of future developments in the LtR domain. We illustrate this by slotting in one of the state-of-the-art LtR objectives - ListMLE~\cite{xia2008listwise} - in place of our pairwise objective (Equation~\ref{eq:pairwiseObj}). As before, we have the output of~\IR~to be $s_{qd}=f(x_{qd})$, and the training of~\IR~is now driven by the optimisation of the ListMLE objective defined as:
\begin{align}
\label{eq:listMLEObj}
\mathcal{L}_{ListMLE} = \sum_{q \in Q} \sum_{I \in \pi(q)} \left[
\sum_{d \in I} s_{qd} - \log\left(\sum_{z \in I:(d \triangleright z)} \exp{(s_{qz})}\right)\right]
\end{align}

We set $t_{\pi}=t_{\mu}=0.5$ and $C=0.1$, and the model architecture for~\IR~is as in the previous section. In Table~\ref{tbl:TrainingObjectives}, the column titled `Swap\%' shows the percentage of training dataset pairs that the trained~\IR~places in an incorrect order. We observe that with the same model, the Pairwise objective fits the training data better - the fraction of swaps induced on the training set by the trained model's scores versus that in the observed rankings is lower. Similarly, the generalisation to unseen queries, represented by the Kendall-$\tau$ of predicted scores with available labels on the validation set, is better with the Pairwise objective. We attribute this to the fact that the ListMLE objective has an implicit bias towards the top of the list to mirror the top-heavy nature of information retrieval metrics. This is at odds with our requirement that the Imitation Ranker effectively reproduce the entire observed rankings.

We make two further observations in this section: (1) Even when using all the features available to the original logging ranker $\pi$, despite being a model of higher capacity (a $2$-layer neural net vs a linear SVM), the imitation ranker is able to achieve a Kendall-$\tau$ of only $0.838$ with a competitive LtR objective - this indicates the difficulty of the task of learning to reproduce entire rankings; (2) Using the ListMLE objective leads to a higher variance estimate - the estimated values for NoC using~\IR~trained via ListMLE was $7.346$, which is significantly higher than the ground-truth ($3.770)$ and what the pairwise RankNet objective yields ($3.713$).

To address ($2$), one option from literature is to truncate the corresponding inverse propensity weight: $min[\frac{1}{\hat{p}(I_k, k | q)}, M]$. A hyper-parameter $M$ has been introduced - setting it to $\infty$ recovers the original estimator. We use $M=100$ and otherwise follow the same process described for IR, and the resulting value of the estimated metrics are indicated by `IR(T)'. We observe how truncation of propensity weights brings the estimated values of NoC and MRR using ListMLE closer to the ground-truth. Note that any value $M<\infty$ would naturally bring down the value of the estimated metric. Therefore, though a truncated version of EP is possible, given that it is already under-estimating the value of the metric, such an option would take it further away from GT. We have therefore not included that combination in the results presented.

We next evaluate models of different complexities and training durations as alternatives for~\IR~which might shed additional light on the behaviour of the two objectives. 

\begin{table}[h!]
  \begin{center}
    \begin{tabular}{|c|c|c|c|c|c|c|}
    \hline
      \multirow{2}{*}{\textbf{Objective}} & \multicolumn{2}{|c|}{\textbf{Ranker Evaluation}} & \multicolumn{4}{|c|}{\textbf{Estimator Evaluation}}\\
      & Swap\% & Kendall-$\tau$ & GT & EP & IR & IR(T)  \\ 
      \hline
      \textbf{Pairwise} & 1.8 & 0.894 & \multirow{2}{*}{3.770} & \multirow{2}{*}{2.151} & 3.713 & 2.916 \\
      \textbf{ListMLE} &  2.7 & 0.838 & & & 7.346 & 3.460 \\
      \bottomrule
    \end{tabular}
    \caption{Comparing the pairwise objective for training the Imitation Ranker with a listwise alternative. The Ranker Evaluation metrics indicate the trained model \IR's ability to reproduce rankings historically produced by $\pi$. The  Estimator Evaluation metrics utilise the trained model's scores in the IPW workflow}
    \label{tbl:TrainingObjectives}
  \end{center}
\end{table}

The task of the IR is a balancing act - it needs to be able to replicate the observed rankings (hence the name), but we would like the propensities derived via the rank distributions to allocate sufficient probability to rare (document,rank) combinations. In the current experiment, we consider variations of the IR model: (a) smaller vs bigger models by varying the number of parameters; and (b) different degrees of learning by varying the number of training epochs. Increases in both (a model of high capacity trained for a long time) will lead to a model that is a better imitator, but will such a model be a better evaluator? 

For the model capacity axis, we have three options: (i) Small: $1$ layer with $700*1$ parameters (ii) Medium: $2$ layers of size $700*32$ and $32*1$ and (iii) Big: $3$ layers corresponding to $700*128$, $128*32$ \& $32*1$ parameters each. In each case, the $700$ corresponds to the number of input features and all non-linearities are tanh. We experiment with numEpochs~$\in \{100, 500, 1000\}$, and show results for both the Pairwise and ListMLE objectives for training \IR. The experiments in earlier sections correspond to the `Medium' model trained for $500$ epochs. 

\begin{figure}[!h]
\centering
\includegraphics[width=\columnwidth]{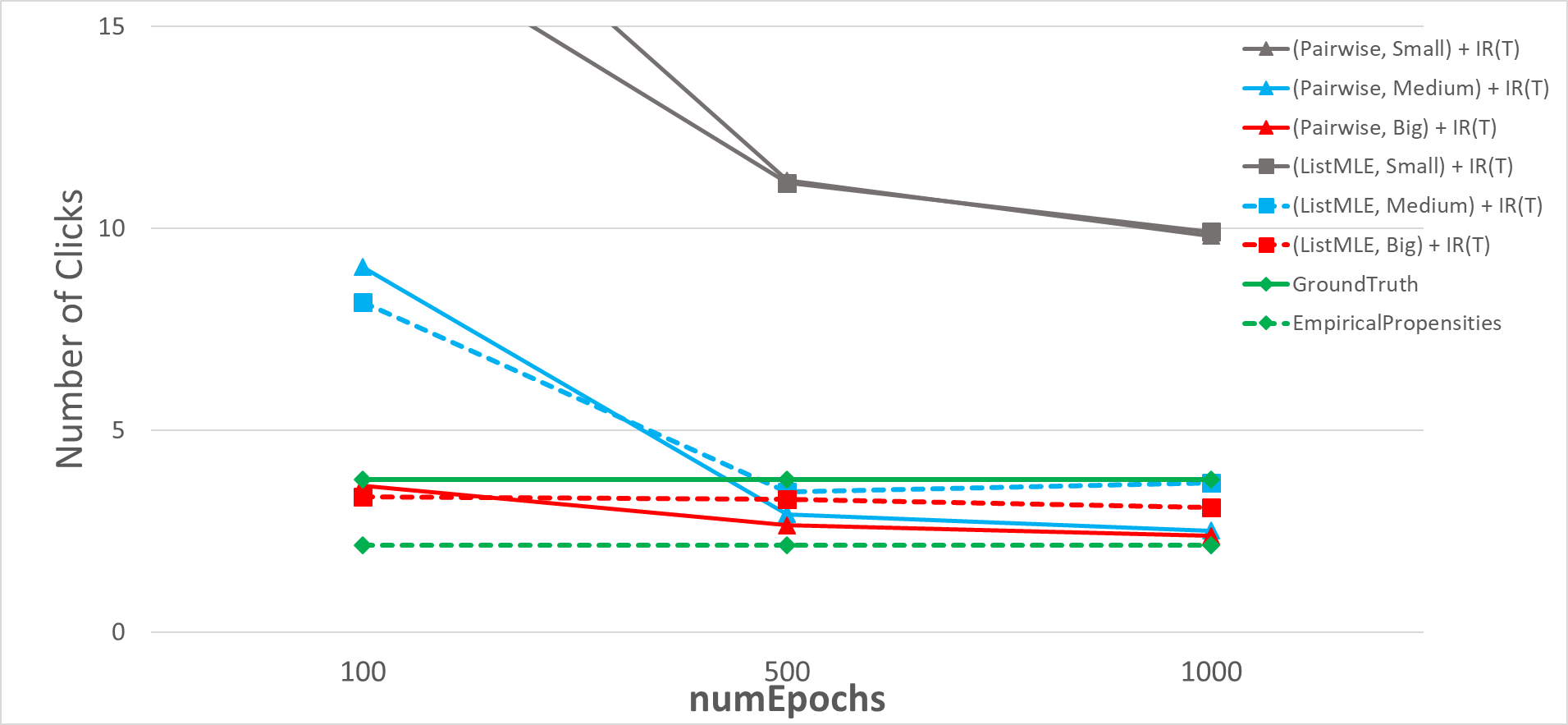}
\caption{NoC estimated by different inferred propensity methods. Swap\% is computed on the training set and Kendall-$\tau$ is computed on the validation set}
\label{fig:modelObjNumEpochsNoC}
\end{figure}

\begin{figure*}[!ht]
\centering
\includegraphics[width=0.85\textwidth,height=0.23\textwidth]{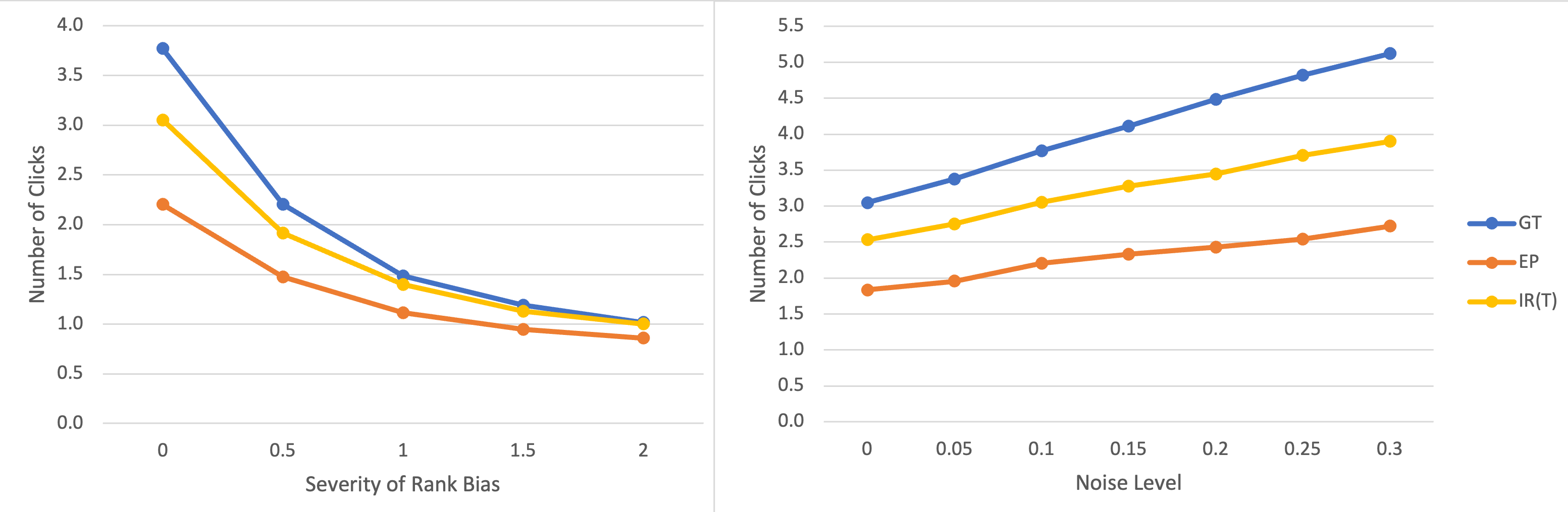}
\caption{Left - The estimated value of NoC as the examination bias ($\eta$) becomes more severe. Right - Behaviour of the different IPW alternatives as noise in the clicks ($\epsilon_{-}$) increases}
\label{fig:paramSensitivity}
\end{figure*}

Figure~\ref{fig:modelObjNumEpochs} provides the ranker metrics for all the combinations. The percentage of swaps induced by the trained ranker's score with respect to the input data is denoted by `Swap\%'. We notice expected behaviour - bigger models trained for longer have lower errors on the training dataset. The Kendall-$\tau$ between rankings induced by the trained ranker scores and the available relevance labels is computed on the validation set. Here we observe typical characteristics of overfitting - for a given objective (Pairwise or ListMLE), the Medium model usually has the best ranking correlation. Comparing the two objectives for a given model size, we notice that ListMLE sometimes has lower error on the train set but the Pairwise objective generalises better. 

In Figure~\ref{fig:modelObjNumEpochsNoC}, the curves closest to the solid green line (representing the Ground Truth) are the most effective combinations. We retain only the version of the IPW estimators with the propensity weights capped at $M=100$. The baseline of using Empirical Propensities is the dashed green line. The Y-axis has been truncated to focus on the region of interest. As a side-effect, the values estimated by the Small model using both the objectives are missing -- the high values for the estimated value of the metric NoC are due to numerically small propensities. The inability of the Small model to perform well highlights the need for a model of sufficient capacity to handle the difficult task of reproducing rankings. It can be noted that the Big model (red solid and dashed lines) does consistently well, even in the setting with fewest number of epochs. 

The method described in the current paper, as well as a majority of the reference literature discussed in Section~\ref{sec:RelatedWork} are based on the inverse propensity weighting (IPW) approach. IPW is used to correct for biases in scenarios when there are factors that affect both the treatment (the choice of document, and the rank at which it was placed) and the outcome (which documents were examined and clicked by the user), and is known to sometimes lead to high variance estimates (as observed in our experiments above). Counterfactual offline evaluation methods based on statistical approaches like the doubly robust estimators have recently been proposed~\cite{kiyohara2022doubly, oosterhuis2022doubly} with the aim of obtaining a better bias-variance trade-off for the evaluation task. Doubly robust methods contain an IPW sub-component, integrating the learnings from our method into the newer work is a subject of future study.

\subsubsection{Parameter Sensitivity Experiments}
\label{sec:ParameterSensitivity}
Our experiments so far simulated clicks on ranked lists produced by $\pi$, these clicks were with respect to a specific user model. In this section, we vary the corresponding parameters so as to evaluate the behaviour of our pipeline under different settings. We first consider the examination bias parameter $\eta$, in the plot on the left in Figure~\ref{fig:paramSensitivity}. The results in earlier sections corresponded to the assumption that the user would examine all ranks equally. While this helped isolate the situation of interest to us - i.e., studying system biases decoupled from user specific aspects - it is an unrealistic assumption in the sense of what we know about how users consume ranked lists. We therefore sweep the presentation bias parameter - leading to reduced likelihood of clicks lower in the ranked list. This setting highlights an interesting observation - existing methods (including list-level matching, and the use of item-level empirical propensities) have an underestimation bias. With fewer clicks overall (towards the right of the X-axis), this essentially leads to the methods looking similar. The use of truncated parametric propensities (i.e., ``IR(T)'') traces the ground-truth most closely. For the plot on the right of Figure~\ref{fig:paramSensitivity}, we progressively increase the click noise parameter $\epsilon_{-}$ which controls the probability of an irrelevant document being clicked. This in general leads to impressions with more clicks, as can be seen by the upward trend for the blue curve representing the ground-truth. The method proposed in the current paper continues to be the best alternative across the range of parameters considered.

\subsubsection{Effect of swaps in the Observation Data} 
For the experiments in the earlier sections, the logged dataset contained impressions in strict order produced by the Observation Ranker $\pi$. In this section, we introduce swaps into the logged dataset to evaluate how sensitive our proposed workflow is to inconsistencies. This experiment is designed to answer research question \textbf{RQ3}, i.e., the swaps are inconsistent with the features available to \IR. 

We follow the process referred to as \textit{RandPair} in~\cite{wang2018position} - from the original impression sorted in decreasing order of score, we swap the results at positions $k$ and $k+1$ with the value of $k$ chosen uniformly between $1$ and $(K-1)$. We introduce a parameter, $B$, to control the fraction of impressions with a swap. Setting $B=0$ recovers the previous setting of all logged rankings being in the order of scores by $\pi$, while $B=50$ means that half the impressions have exactly one swap. The simulated logged dataset, with the swaps introduced, are input into the regular pipeline of training the Imitation Ranker before producing the rank distributions to be used as inverse propensity weights. Other experimental parameters are retained from previous experiments: $t_{\pi}=t_{\mu}=0.5$, $\eta=0$, $\epsilon_{-}$ and $C=0.1$, the Imitation Ranker model corresponds to Medium and the truncation constant $M=100$. 

\begin{table}[!ht]
\label{tbl:LengthOfBGroupSweep}
\centering
\begin{tabular}{|l|c|c|c|c|c|c|} 
\hline
\multirow{2}{*}{B} & 
\multirow{2}{*}{GT} &
\multirow{2}{*}{EP} &
\multicolumn{2}{c|}{Pairwise} &
\multicolumn{2}{c|}{ListMLE} \\
 & & & IR & IR(T) & IR & IR(T) \\ 
\hline
  0 & 3.770 & 2.151 &   3.713 & 2.916 &   7.346 & 3.460 \\
 10 & 3.752 & 2.332 &  51.695 & 3.116 &  12.819 & 3.520 \\
 20 & 3.757 & 2.477 &  95.048 & 3.280 &  30.622 & 3.500 \\
 30 & 3.756 & 2.584 & 157.923 & 3.618 &  69.693 & 3.820 \\
 40 & 3.766 & 2.672 & 163.502 & 3.568 &  90.722 & 3.655 \\
 50 & 3.763 & 2.675 & 248.172 & 3.721 &  94.858 & 3.631 \\
 60 & 3.771 & 2.789 & 310.939 & 4.007 & 108.952 & 3.788 \\
 70 & 3.774 & 2.835 & 514.044 & 4.207 & 162.179 & 3.984 \\
 80 & 3.772 & 2.799 & 528.003 & 4.296 & 236.164 & 4.133 \\
 90 & 3.764 & 2.841 & 628.854 & 4.482 & 225.261 & 4.141 \\
100 & 3.769 & 2.841 & 609.708 & 4.792 & 320.920 & 4.319 \\
\bottomrule
\end{tabular}
\caption{When observed rankings are consistent with available features, Pairwise performs better than ListMLE. In the presence of swaps, propensities obtained via ListMLE are more robust. With truncation, both options approach the GT}
\end{table}

We look at how the propensities produced by the models affect the estimation of the number of clicks for impressions in the Measurement Set. While not displayed here, the trends for MRR are the same. By design, the Ground Truth changes very marginally with increasing swaps -- our simulated clicks are a function of the relevance labels alone, not the ranks. In general, the propensity of a matched (document,rank) pair reduces as the input data has more swaps, this manifests as an increasing trend in the IPW estimate using Empirical Propensities. However, the propensities output by the parameterised workflow proposed here exhibit a much more catastrophic behaviour. We have touched upon this high variance characteristic in previous sections. And as there, we note that truncation (using a constant value that we have not tuned) produces estimates that are in the correct range. While the imitation ranker with truncated propensities (represented by IR(T)) is consistently the best option across all levels of swap, this experiment illustrates the need to more carefully train~\IR~so that it is more robust with respect to possible inconsistencies. In particular, the current experiment suggests that the use of IPW based evaluation using estimated propensities needs the features $x_{qd}$ available to the IR to be complete with respect to the rankings logged in $D_\pi$.

\section{Discussion and Limitations}

In this paper, we considered the problem of offline evaluation of ranked lists from logged clickthrough data. We focused on the effect of differences between the ranker that generated historical impressions and the new ranker to be evaluated. These differences are with respect to the documents they place at specific ranks - a bias that we account for in an Inverse Propensity Weighting based formulation. 

The novel contribution is the proposal of a `Learning to Evaluate' workflow where the ``learning'' component refers to the building of an Imitation Ranker (\IR) that is trained to mimic the behaviour of the historical ranker. Since the~\IR~uses features, we can derive smoother (document,rank) propensities than we would if they were derived empirically from the data. The smoothed propensities are what enable the IPW estimator to produce reasonable evaluations in the presence of differences between the rankers. We also made the observation that if the observed rankings are not explained by the features available to~\IR, the resulting low propensities lead to a high variance evaluation. Future work will look into ensuring that~\IR~allocates sufficient probability mass to rare (document,rank) pairs.

The final propensities generated by the IR are a combination of multiple factors - the available features, the model, as well as the training method. Our experiments illustrated what are interesting differences between a standard LtR task versus that of producing~\IR~to be used in an evaluation setting. The main observation was that unlike standard top-heavy LtR objectives, the training of~\IR~needs to focus on all positions in the rankings. Our experimental section suggests that in addition to the objective, we require high capacity models to reproduce the observed rankings.

While we have conducted an extensive study on many factors, few others remain. For e.g., integrating our investigations of the effect of system biases into existing studies on user-centric biases will lead to a more comprehensive treatment. In addition, validating the offline estimates of the relevance of the new ranker via online A/B testing would provide additional support to the method described here.

\section{Conclusions and Future Work}

Test collection based offline evaluation of IR systems is a very well studied topic~\cite{sandersonBook}. Reusability~\cite{carterette2010measuring} is an important motivation while constructing reference test collections. That is, confidence in an evaluation protocol requires the ability to reliably evaluate rankings different from those used to obtain relevance labels. There are at least two critical components of a reliable evaluation: (a) the availability of an exhaustive set of judgements; and (b) metrics that are robust to the presence of incomplete information~\cite{buckley2004}. In the context of TREC, the practice of pooling the results of a range of diverse systems is an attempt to achieve (a). Though pooling has its limitations~\cite{buckley2007}, it is worth noting that observed clicks in observational datasets from a production rankers will likely not have the benefits of a range of diverse top-$K$ lists having been examined by users. 

The experimental results provide two avenues of future work: (a) In general, the proposed workflow accounts for the bias induced by the specific ranking shown by the previous ranker. This continues to be a concern for traditional judgements based evaluation~\cite{yilmaz2020reliability}, we wish to investigate what aspects of the method proposed here transfer to that setting (b) While the focus in the current paper is on \textit{evaluation}, offline \textit{learning} from biased data is a natural next step.

\bibliographystyle{ACM-Reference-Format}
\balance
\bibliography{references}


\begin{thebibliography}{48}


\ifx \showCODEN    \undefined \def \showCODEN     #1{\unskip}     \fi
\ifx \showDOI      \undefined \def \showDOI       #1{#1}\fi
\ifx \showISBNx    \undefined \def \showISBNx     #1{\unskip}     \fi
\ifx \showISBNxiii \undefined \def \showISBNxiii  #1{\unskip}     \fi
\ifx \showISSN     \undefined \def \showISSN      #1{\unskip}     \fi
\ifx \showLCCN     \undefined \def \showLCCN      #1{\unskip}     \fi
\ifx \shownote     \undefined \def \shownote      #1{#1}          \fi
\ifx \showarticletitle \undefined \def \showarticletitle #1{#1}   \fi
\ifx \showURL      \undefined \def \showURL       {\relax}        \fi
\providecommand\bibfield[2]{#2}
\providecommand\bibinfo[2]{#2}
\providecommand\natexlab[1]{#1}
\providecommand\showeprint[2][]{arXiv:#2}

\bibitem[Agarwal et~al\mbox{.}(2019a)]%
        {agarwal2019general}
\bibfield{author}{\bibinfo{person}{Aman Agarwal}, \bibinfo{person}{Kenta
  Takatsu}, \bibinfo{person}{Ivan Zaitsev}, {and} \bibinfo{person}{Thorsten
  Joachims}.} \bibinfo{year}{2019}\natexlab{a}.
\newblock \showarticletitle{A general framework for counterfactual
  learning-to-rank}. In \bibinfo{booktitle}{\emph{Proceedings of the 42nd
  International ACM SIGIR Conference on Research and Development in Information
  Retrieval}}. \bibinfo{pages}{5--14}.
\newblock


\bibitem[Agarwal et~al\mbox{.}(2019b)]%
        {agarwal2019addressing}
\bibfield{author}{\bibinfo{person}{Aman Agarwal}, \bibinfo{person}{Xuanhui
  Wang}, \bibinfo{person}{Cheng Li}, \bibinfo{person}{Michael Bendersky}, {and}
  \bibinfo{person}{Marc Najork}.} \bibinfo{year}{2019}\natexlab{b}.
\newblock \showarticletitle{Addressing trust bias for unbiased
  learning-to-rank}. In \bibinfo{booktitle}{\emph{The World Wide Web
  Conference}}. \bibinfo{pages}{4--14}.
\newblock


\bibitem[Agarwal et~al\mbox{.}(2018)]%
        {agarwal2018consistent}
\bibfield{author}{\bibinfo{person}{Aman Agarwal}, \bibinfo{person}{Ivan
  Zaitsev}, {and} \bibinfo{person}{Thorsten Joachims}.}
  \bibinfo{year}{2018}\natexlab{}.
\newblock \showarticletitle{Consistent position bias estimation without online
  interventions for learning-to-rank}.
\newblock \bibinfo{journal}{\emph{arXiv preprint arXiv:1806.03555}}
  (\bibinfo{year}{2018}).
\newblock


\bibitem[Agarwal et~al\mbox{.}(2019c)]%
        {agarwal2019estimating}
\bibfield{author}{\bibinfo{person}{Aman Agarwal}, \bibinfo{person}{Ivan
  Zaitsev}, \bibinfo{person}{Xuanhui Wang}, \bibinfo{person}{Cheng Li},
  \bibinfo{person}{Marc Najork}, {and} \bibinfo{person}{Thorsten Joachims}.}
  \bibinfo{year}{2019}\natexlab{c}.
\newblock \showarticletitle{Estimating position bias without intrusive
  interventions}. In \bibinfo{booktitle}{\emph{Proceedings of the Twelfth ACM
  International Conference on Web Search and Data Mining}}.
  \bibinfo{pages}{474--482}.
\newblock


\bibitem[Ai et~al\mbox{.}(2018)]%
        {ai2018unbiased}
\bibfield{author}{\bibinfo{person}{Qingyao Ai}, \bibinfo{person}{Keping Bi},
  \bibinfo{person}{Cheng Luo}, \bibinfo{person}{Jiafeng Guo}, {and}
  \bibinfo{person}{W~Bruce Croft}.} \bibinfo{year}{2018}\natexlab{}.
\newblock \showarticletitle{Unbiased learning to rank with unbiased propensity
  estimation}. In \bibinfo{booktitle}{\emph{The 41st International ACM SIGIR
  Conference on Research \& Development in Information Retrieval}}.
  \bibinfo{pages}{385--394}.
\newblock


\bibitem[Breuer et~al\mbox{.}(2020)]%
        {breuer2020measure}
\bibfield{author}{\bibinfo{person}{Timo Breuer}, \bibinfo{person}{Nicola
  Ferro}, \bibinfo{person}{Norbert Fuhr}, \bibinfo{person}{Maria Maistro},
  \bibinfo{person}{Tetsuya Sakai}, \bibinfo{person}{Philipp Schaer}, {and}
  \bibinfo{person}{Ian Soboroff}.} \bibinfo{year}{2020}\natexlab{}.
\newblock \showarticletitle{How to measure the reproducibility of
  system-oriented IR experiments}. In \bibinfo{booktitle}{\emph{Proceedings of
  the 43rd International ACM SIGIR Conference on Research and Development in
  Information Retrieval}}. \bibinfo{pages}{349--358}.
\newblock


\bibitem[Buckley et~al\mbox{.}(2007)]%
        {buckley2007}
\bibfield{author}{\bibinfo{person}{Chris Buckley}, \bibinfo{person}{Darrin
  Dimmick}, \bibinfo{person}{Ian Soboroff}, {and} \bibinfo{person}{Ellen
  Voorhees}.} \bibinfo{year}{2007}\natexlab{}.
\newblock \showarticletitle{Bias and the limits of pooling for large
  collections}.
\newblock \bibinfo{journal}{\emph{Information retrieval}} \bibinfo{volume}{10},
  \bibinfo{number}{6} (\bibinfo{year}{2007}), \bibinfo{pages}{491--508}.
\newblock


\bibitem[Buckley and Voorhees(2004)]%
        {buckley2004}
\bibfield{author}{\bibinfo{person}{Chris Buckley} {and}
  \bibinfo{person}{Ellen~M Voorhees}.} \bibinfo{year}{2004}\natexlab{}.
\newblock \showarticletitle{Retrieval evaluation with incomplete information}.
  In \bibinfo{booktitle}{\emph{Proceedings of the 27th annual international ACM
  SIGIR conference on Research and development in information retrieval}}.
  \bibinfo{pages}{25--32}.
\newblock


\bibitem[Burges et~al\mbox{.}(2005)]%
        {burges2005learning}
\bibfield{author}{\bibinfo{person}{Chris Burges}, \bibinfo{person}{Tal Shaked},
  \bibinfo{person}{Erin Renshaw}, \bibinfo{person}{Ari Lazier},
  \bibinfo{person}{Matt Deeds}, \bibinfo{person}{Nicole Hamilton}, {and}
  \bibinfo{person}{Greg Hullender}.} \bibinfo{year}{2005}\natexlab{}.
\newblock \showarticletitle{Learning to rank using gradient descent}. In
  \bibinfo{booktitle}{\emph{Proceedings of the 22nd international conference on
  Machine learning}}. \bibinfo{pages}{89--96}.
\newblock


\bibitem[Carterette(2011)]%
        {carterette2011}
\bibfield{author}{\bibinfo{person}{Ben Carterette}.}
  \bibinfo{year}{2011}\natexlab{}.
\newblock \showarticletitle{System effectiveness, user models, and user
  utility: a conceptual framework for investigation}. In
  \bibinfo{booktitle}{\emph{Proceedings of the 34th international ACM SIGIR
  conference on Research and development in Information Retrieval}}.
  \bibinfo{pages}{903--912}.
\newblock


\bibitem[Carterette et~al\mbox{.}(2010)]%
        {carterette2010measuring}
\bibfield{author}{\bibinfo{person}{Ben Carterette}, \bibinfo{person}{Evgeniy
  Gabrilovich}, \bibinfo{person}{Vanja Josifovski}, {and}
  \bibinfo{person}{Donald Metzler}.} \bibinfo{year}{2010}\natexlab{}.
\newblock \showarticletitle{Measuring the reusability of test collections}. In
  \bibinfo{booktitle}{\emph{Proceedings of the third ACM international
  conference on Web search and data mining}}. \bibinfo{pages}{231--240}.
\newblock


\bibitem[Chuklin et~al\mbox{.}(2015)]%
        {chuklin2015click}
\bibfield{author}{\bibinfo{person}{Aleksandr Chuklin}, \bibinfo{person}{Ilya
  Markov}, {and} \bibinfo{person}{Maarten~de Rijke}.}
  \bibinfo{year}{2015}\natexlab{}.
\newblock \showarticletitle{Click models for web search}.
\newblock \bibinfo{journal}{\emph{Synthesis lectures on information concepts,
  retrieval, and services}} \bibinfo{volume}{7}, \bibinfo{number}{3}
  (\bibinfo{year}{2015}), \bibinfo{pages}{1--115}.
\newblock


\bibitem[Craswell et~al\mbox{.}(2008)]%
        {craswell2008experimental}
\bibfield{author}{\bibinfo{person}{Nick Craswell}, \bibinfo{person}{Onno
  Zoeter}, \bibinfo{person}{Michael Taylor}, {and} \bibinfo{person}{Bill
  Ramsey}.} \bibinfo{year}{2008}\natexlab{}.
\newblock \showarticletitle{An experimental comparison of click position-bias
  models}. In \bibinfo{booktitle}{\emph{Proceedings of the 2008 international
  conference on web search and data mining}}. \bibinfo{pages}{87--94}.
\newblock


\bibitem[Dehghani et~al\mbox{.}(2017)]%
        {dehghani2017neural}
\bibfield{author}{\bibinfo{person}{Mostafa Dehghani}, \bibinfo{person}{Hamed
  Zamani}, \bibinfo{person}{Aliaksei Severyn}, \bibinfo{person}{Jaap Kamps},
  {and} \bibinfo{person}{W~Bruce Croft}.} \bibinfo{year}{2017}\natexlab{}.
\newblock \showarticletitle{Neural ranking models with weak supervision}. In
  \bibinfo{booktitle}{\emph{Proceedings of the 40th International ACM SIGIR
  Conference on Research and Development in Information Retrieval}}.
  \bibinfo{pages}{65--74}.
\newblock


\bibitem[Diaz et~al\mbox{.}(2020)]%
        {diaz2020evaluating}
\bibfield{author}{\bibinfo{person}{Fernando Diaz}, \bibinfo{person}{Bhaskar
  Mitra}, \bibinfo{person}{Michael~D Ekstrand}, \bibinfo{person}{Asia~J Biega},
  {and} \bibinfo{person}{Ben Carterette}.} \bibinfo{year}{2020}\natexlab{}.
\newblock \showarticletitle{Evaluating Stochastic Rankings with Expected
  Exposure}.
\newblock \bibinfo{journal}{\emph{arXiv preprint arXiv:2004.13157}}
  (\bibinfo{year}{2020}).
\newblock


\bibitem[Fang et~al\mbox{.}(2019)]%
        {fang2019intervention}
\bibfield{author}{\bibinfo{person}{Zhichong Fang}, \bibinfo{person}{Aman
  Agarwal}, {and} \bibinfo{person}{Thorsten Joachims}.}
  \bibinfo{year}{2019}\natexlab{}.
\newblock \showarticletitle{Intervention harvesting for context-dependent
  examination-bias estimation}. In \bibinfo{booktitle}{\emph{Proceedings of the
  42nd International ACM SIGIR Conference on Research and Development in
  Information Retrieval}}. \bibinfo{pages}{825--834}.
\newblock


\bibitem[Gilotte et~al\mbox{.}(2018)]%
        {ref_lncs10}
\bibfield{author}{\bibinfo{person}{Alexandre Gilotte},
  \bibinfo{person}{Cl{\'e}ment Calauz{\`e}nes}, \bibinfo{person}{Thomas
  Nedelec}, \bibinfo{person}{Alexandre Abraham}, {and} \bibinfo{person}{Simon
  Doll{\'e}}.} \bibinfo{year}{2018}\natexlab{}.
\newblock \showarticletitle{Offline a/b testing for recommender systems}. In
  \bibinfo{booktitle}{\emph{Proceedings of the Eleventh ACM International
  Conference on Web Search and Data Mining}}. \bibinfo{pages}{198--206}.
\newblock


\bibitem[Hinton et~al\mbox{.}(2015)]%
        {hinton2015distilling}
\bibfield{author}{\bibinfo{person}{Geoffrey Hinton}, \bibinfo{person}{Oriol
  Vinyals}, {and} \bibinfo{person}{Jeff Dean}.}
  \bibinfo{year}{2015}\natexlab{}.
\newblock \showarticletitle{Distilling the knowledge in a neural network}.
\newblock \bibinfo{journal}{\emph{arXiv preprint arXiv:1503.02531}}
  (\bibinfo{year}{2015}).
\newblock


\bibitem[Hofmann et~al\mbox{.}(2016)]%
        {onlineEvalIR}
\bibfield{author}{\bibinfo{person}{Katja Hofmann}, \bibinfo{person}{Lihong Li},
  {and} \bibinfo{person}{Filip Radlinski}.} \bibinfo{year}{2016}\natexlab{}.
\newblock \showarticletitle{Online evaluation for information retrieval}.
\newblock \bibinfo{journal}{\emph{Foundations and Trends in Information
  Retrieval}} \bibinfo{volume}{10}, \bibinfo{number}{1} (\bibinfo{year}{2016}),
  \bibinfo{pages}{1--117}.
\newblock


\bibitem[Hu et~al\mbox{.}(2018)]%
        {lambdaMart}
\bibfield{author}{\bibinfo{person}{Ziniu Hu}, \bibinfo{person}{Yang Wang},
  \bibinfo{person}{Qu Peng}, {and} \bibinfo{person}{Hang Li}.}
  \bibinfo{year}{2018}\natexlab{}.
\newblock \showarticletitle{A Novel Algorithm for Unbiased Learning to Rank}.
\newblock \bibinfo{journal}{\emph{CoRR}}  \bibinfo{volume}{abs/1809.05818}
  (\bibinfo{year}{2018}).
\newblock
\showeprint[arxiv]{1809.05818}
\urldef\tempurl%
\url{http://arxiv.org/abs/1809.05818}
\showURL{%
\tempurl}


\bibitem[Joachims and Swaminathan(2016)]%
        {sigir2016Tutorial}
\bibfield{author}{\bibinfo{person}{Thorsten Joachims} {and}
  \bibinfo{person}{Adith Swaminathan}.} \bibinfo{year}{2016}\natexlab{}.
\newblock \showarticletitle{Counterfactual evaluation and learning for search,
  recommendation and ad placement}. In \bibinfo{booktitle}{\emph{Proceedings of
  the 39th International ACM SIGIR conference on Research and Development in
  Information Retrieval}}. \bibinfo{pages}{1199--1201}.
\newblock


\bibitem[Joachims et~al\mbox{.}(2017)]%
        {joachims2017}
\bibfield{author}{\bibinfo{person}{Thorsten Joachims}, \bibinfo{person}{Adith
  Swaminathan}, {and} \bibinfo{person}{Tobias Schnabel}.}
  \bibinfo{year}{2017}\natexlab{}.
\newblock \showarticletitle{Unbiased learning-to-rank with biased feedback}. In
  \bibinfo{booktitle}{\emph{Proceedings of the Tenth ACM International
  Conference on Web Search and Data Mining}}. \bibinfo{pages}{781--789}.
\newblock


\bibitem[Kiyohara et~al\mbox{.}(2022)]%
        {kiyohara2022doubly}
\bibfield{author}{\bibinfo{person}{Haruka Kiyohara}, \bibinfo{person}{Yuta
  Saito}, \bibinfo{person}{Tatsuya Matsuhiro}, \bibinfo{person}{Yusuke Narita},
  \bibinfo{person}{Nobuyuki Shimizu}, {and} \bibinfo{person}{Yasuo Yamamoto}.}
  \bibinfo{year}{2022}\natexlab{}.
\newblock \showarticletitle{Doubly Robust Off-Policy Evaluation for Ranking
  Policies under the Cascade Behavior Model}. In
  \bibinfo{booktitle}{\emph{Proceedings of the Fifteenth ACM International
  Conference on Web Search and Data Mining}}. \bibinfo{pages}{487--497}.
\newblock


\bibitem[Kohavi et~al\mbox{.}(2013)]%
        {kohavi2013}
\bibfield{author}{\bibinfo{person}{Ron Kohavi}, \bibinfo{person}{Alex Deng},
  \bibinfo{person}{Brian Frasca}, \bibinfo{person}{Toby Walker},
  \bibinfo{person}{Ya Xu}, {and} \bibinfo{person}{Nils Pohlmann}.}
  \bibinfo{year}{2013}\natexlab{}.
\newblock \showarticletitle{Online controlled experiments at large scale}. In
  \bibinfo{booktitle}{\emph{Proceedings of the 19th ACM SIGKDD international
  conference on Knowledge discovery and data mining}}.
  \bibinfo{pages}{1168--1176}.
\newblock


\bibitem[Li(2015)]%
        {wsdm2015Tutorial}
\bibfield{author}{\bibinfo{person}{Lihong Li}.}
  \bibinfo{year}{2015}\natexlab{}.
\newblock \showarticletitle{Offline evaluation and optimization for interactive
  systems}.
\newblock  (\bibinfo{year}{2015}).
\newblock


\bibitem[Li et~al\mbox{.}(2010)]%
        {replay}
\bibfield{author}{\bibinfo{person}{Lihong Li}, \bibinfo{person}{Wei Chu}, {and}
  \bibinfo{person}{John Langford}.} \bibinfo{year}{2010}\natexlab{}.
\newblock \showarticletitle{An Unbiased, Data-Driven, Offline Evaluation Method
  of Contextual Bandit Algorithms}.
\newblock \bibinfo{journal}{\emph{CoRR}}  \bibinfo{volume}{abs/1003.5956}
  (\bibinfo{year}{2010}).
\newblock
\showeprint[arxiv]{1003.5956}
\urldef\tempurl%
\url{http://arxiv.org/abs/1003.5956}
\showURL{%
\tempurl}


\bibitem[Li et~al\mbox{.}(2015)]%
        {lihong2015}
\bibfield{author}{\bibinfo{person}{Lihong Li}, \bibinfo{person}{Jin~Young Kim},
  {and} \bibinfo{person}{Imed Zitouni}.} \bibinfo{year}{2015}\natexlab{}.
\newblock \showarticletitle{Toward predicting the outcome of an A/B experiment
  for search relevance}. In \bibinfo{booktitle}{\emph{Proceedings of the Eighth
  ACM International Conference on Web Search and Data Mining}}.
  \bibinfo{pages}{37--46}.
\newblock


\bibitem[Li et~al\mbox{.}(2018)]%
        {li2018offline}
\bibfield{author}{\bibinfo{person}{Shuai Li}, \bibinfo{person}{Yasin
  Abbasi-Yadkori}, \bibinfo{person}{Branislav Kveton}, \bibinfo{person}{S
  Muthukrishnan}, \bibinfo{person}{Vishwa Vinay}, {and} \bibinfo{person}{Zheng
  Wen}.} \bibinfo{year}{2018}\natexlab{}.
\newblock \showarticletitle{Offline evaluation of ranking policies with click
  models}. In \bibinfo{booktitle}{\emph{Proceedings of the 24th ACM SIGKDD
  International Conference on Knowledge Discovery \& Data Mining}}.
  \bibinfo{pages}{1685--1694}.
\newblock


\bibitem[Lioma et~al\mbox{.}(2017)]%
        {lioma17}
\bibfield{author}{\bibinfo{person}{Christina Lioma},
  \bibinfo{person}{Jakob~Grue Simonsen}, {and} \bibinfo{person}{Birger
  Larsen}.} \bibinfo{year}{2017}\natexlab{}.
\newblock \showarticletitle{Evaluation measures for relevance and credibility
  in ranked lists}. In \bibinfo{booktitle}{\emph{Proceedings of the ACM SIGIR
  International Conference on Theory of Information Retrieval}}.
  \bibinfo{pages}{91--98}.
\newblock


\bibitem[Lu et~al\mbox{.}(2017)]%
        {lu2017}
\bibfield{author}{\bibinfo{person}{Xiaolu Lu}, \bibinfo{person}{Alistair
  Moffat}, {and} \bibinfo{person}{J~Shane Culpepper}.}
  \bibinfo{year}{2017}\natexlab{}.
\newblock \showarticletitle{Can Deep Effectiveness Metrics Be Evaluated Using
  Shallow Judgment Pools?}. In \bibinfo{booktitle}{\emph{Proceedings of the
  40th International ACM SIGIR Conference on Research and Development in
  Information Retrieval}}. \bibinfo{pages}{35--44}.
\newblock


\bibitem[Oosterhuis(2022)]%
        {oosterhuis2022doubly}
\bibfield{author}{\bibinfo{person}{Harrie Oosterhuis}.}
  \bibinfo{year}{2022}\natexlab{}.
\newblock \showarticletitle{Doubly-Robust Estimation for Unbiased
  Learning-to-Rank from Position-Biased Click Feedback}.
\newblock \bibinfo{journal}{\emph{arXiv preprint arXiv:2203.17118}}
  (\bibinfo{year}{2022}).
\newblock


\bibitem[Oosterhuis and de~Rijke(2018)]%
        {oosterhuis2018differentiable}
\bibfield{author}{\bibinfo{person}{Harrie Oosterhuis} {and}
  \bibinfo{person}{Maarten de Rijke}.} \bibinfo{year}{2018}\natexlab{}.
\newblock \showarticletitle{Differentiable unbiased online learning to rank}.
  In \bibinfo{booktitle}{\emph{Proceedings of the 27th ACM International
  Conference on Information and Knowledge Management}}.
  \bibinfo{pages}{1293--1302}.
\newblock


\bibitem[Oosterhuis and de~Rijke(2021)]%
        {oosterhuis2021unifying}
\bibfield{author}{\bibinfo{person}{Harrie Oosterhuis} {and}
  \bibinfo{person}{Maarten de Rijke}.} \bibinfo{year}{2021}\natexlab{}.
\newblock \showarticletitle{Unifying online and counterfactual learning to
  rank: A novel counterfactual estimator that effectively utilizes online
  interventions}. In \bibinfo{booktitle}{\emph{Proceedings of the 14th ACM
  International Conference on Web Search and Data Mining}}.
  \bibinfo{pages}{463--471}.
\newblock


\bibitem[Oosterhuis et~al\mbox{.}(2019)]%
        {sigir2019Tutorial}
\bibfield{author}{\bibinfo{person}{Harrie Oosterhuis}, \bibinfo{person}{Rolf
  Jagerman}, {and} \bibinfo{person}{Maarten de Rijke}.}
  \bibinfo{year}{2019}\natexlab{}.
\newblock \showarticletitle{Unbiased Learning to Rank: Counterfactual and
  Online Approaches}.
\newblock \bibinfo{journal}{\emph{CoRR}}  \bibinfo{volume}{abs/1907.07260}
  (\bibinfo{year}{2019}).
\newblock
\showeprint[arxiv]{1907.07260}
\urldef\tempurl%
\url{http://arxiv.org/abs/1907.07260}
\showURL{%
\tempurl}


\bibitem[Oosterhuis et~al\mbox{.}(2020)]%
        {oosterhuis2020unbiased}
\bibfield{author}{\bibinfo{person}{Harrie Oosterhuis}, \bibinfo{person}{Rolf
  Jagerman}, {and} \bibinfo{person}{Maarten de Rijke}.}
  \bibinfo{year}{2020}\natexlab{}.
\newblock \showarticletitle{Unbiased Learning to Rank: Counterfactual and
  Online Approaches}. In \bibinfo{booktitle}{\emph{Companion Proceedings of the
  Web Conference 2020}}. \bibinfo{pages}{299--300}.
\newblock


\bibitem[Ovaisi et~al\mbox{.}(2020)]%
        {ovaisi2020correcting}
\bibfield{author}{\bibinfo{person}{Zohreh Ovaisi}, \bibinfo{person}{Ragib
  Ahsan}, \bibinfo{person}{Yifan Zhang}, \bibinfo{person}{Kathryn Vasilaky},
  {and} \bibinfo{person}{Elena Zheleva}.} \bibinfo{year}{2020}\natexlab{}.
\newblock \showarticletitle{Correcting for Selection Bias in Learning-to-rank
  Systems}. In \bibinfo{booktitle}{\emph{Proceedings of The Web Conference
  2020}}. \bibinfo{pages}{1863--1873}.
\newblock


\bibitem[Ovaisi et~al\mbox{.}(2021)]%
        {ovaisi2021propensity}
\bibfield{author}{\bibinfo{person}{Zohreh Ovaisi}, \bibinfo{person}{Kathryn
  Vasilaky}, {and} \bibinfo{person}{Elena Zheleva}.}
  \bibinfo{year}{2021}\natexlab{}.
\newblock \showarticletitle{Propensity-Independent Bias Recovery in Offline
  Learning-to-Rank Systems}. In \bibinfo{booktitle}{\emph{Proceedings of the
  44th International ACM SIGIR Conference on Research and Development in
  Information Retrieval}}. \bibinfo{pages}{1763--1767}.
\newblock


\bibitem[Rosenbaum and Rubin(1983)]%
        {ref_lncs5}
\bibfield{author}{\bibinfo{person}{Paul~R Rosenbaum} {and}
  \bibinfo{person}{Donald~B Rubin}.} \bibinfo{year}{1983}\natexlab{}.
\newblock \showarticletitle{The central role of the propensity score in
  observational studies for causal effects}.
\newblock \bibinfo{journal}{\emph{Biometrika}} \bibinfo{volume}{70},
  \bibinfo{number}{1} (\bibinfo{year}{1983}), \bibinfo{pages}{41--55}.
\newblock


\bibitem[Sanderson(2010)]%
        {sandersonBook}
\bibfield{author}{\bibinfo{person}{Mark Sanderson}.}
  \bibinfo{year}{2010}\natexlab{}.
\newblock \bibinfo{booktitle}{\emph{Test collection based evaluation of
  information retrieval systems}}.
\newblock \bibinfo{publisher}{Now Publishers Inc}.
\newblock


\bibitem[Schnabel et~al\mbox{.}(2016)]%
        {ref_lncs1}
\bibfield{author}{\bibinfo{person}{Tobias Schnabel}, \bibinfo{person}{Adith
  Swaminathan}, \bibinfo{person}{Peter~I Frazier}, {and}
  \bibinfo{person}{Thorsten Joachims}.} \bibinfo{year}{2016}\natexlab{}.
\newblock \showarticletitle{Unbiased comparative evaluation of ranking
  functions}. In \bibinfo{booktitle}{\emph{Proceedings of the 2016 ACM
  International Conference on the Theory of Information Retrieval}}.
  \bibinfo{pages}{109--118}.
\newblock


\bibitem[Schuth et~al\mbox{.}(2014)]%
        {schuth2014}
\bibfield{author}{\bibinfo{person}{Anne Schuth}, \bibinfo{person}{Floor
  Sietsma}, \bibinfo{person}{Shimon Whiteson}, \bibinfo{person}{Damien
  Lefortier}, {and} \bibinfo{person}{Maarten de Rijke}.}
  \bibinfo{year}{2014}\natexlab{}.
\newblock \showarticletitle{Multileaved comparisons for fast online
  evaluation}. In \bibinfo{booktitle}{\emph{Proceedings of the 23rd ACM
  International Conference on Conference on Information and Knowledge
  Management}}. \bibinfo{pages}{71--80}.
\newblock


\bibitem[Siroker and Koomen(2013)]%
        {abTesting}
\bibfield{author}{\bibinfo{person}{Dan Siroker} {and} \bibinfo{person}{Pete
  Koomen}.} \bibinfo{year}{2013}\natexlab{}.
\newblock \bibinfo{booktitle}{\emph{A/B testing: The most powerful way to turn
  clicks into customers}}.
\newblock \bibinfo{publisher}{John Wiley \& Sons}.
\newblock


\bibitem[Taylor et~al\mbox{.}(2008)]%
        {taylor2008softrank}
\bibfield{author}{\bibinfo{person}{Michael Taylor}, \bibinfo{person}{John
  Guiver}, \bibinfo{person}{Stephen Robertson}, {and} \bibinfo{person}{Tom
  Minka}.} \bibinfo{year}{2008}\natexlab{}.
\newblock \showarticletitle{SoftRank: optimizing non-smooth rank metrics}. In
  \bibinfo{booktitle}{\emph{Proceedings of the 2008 International Conference on
  Web Search and Data Mining}}. \bibinfo{pages}{77--86}.
\newblock


\bibitem[Wang et~al\mbox{.}(2016)]%
        {wang2016}
\bibfield{author}{\bibinfo{person}{Xuanhui Wang}, \bibinfo{person}{Michael
  Bendersky}, \bibinfo{person}{Donald Metzler}, {and} \bibinfo{person}{Marc
  Najork}.} \bibinfo{year}{2016}\natexlab{}.
\newblock \showarticletitle{Learning to rank with selection bias in personal
  search}. In \bibinfo{booktitle}{\emph{Proceedings of the 39th International
  ACM SIGIR conference on Research and Development in Information Retrieval}}.
  \bibinfo{pages}{115--124}.
\newblock


\bibitem[Wang et~al\mbox{.}(2018)]%
        {wang2018position}
\bibfield{author}{\bibinfo{person}{Xuanhui Wang}, \bibinfo{person}{Nadav
  Golbandi}, \bibinfo{person}{Michael Bendersky}, \bibinfo{person}{Donald
  Metzler}, {and} \bibinfo{person}{Marc Najork}.}
  \bibinfo{year}{2018}\natexlab{}.
\newblock \showarticletitle{Position bias estimation for unbiased learning to
  rank in personal search}. In \bibinfo{booktitle}{\emph{Proceedings of the
  Eleventh ACM International Conference on Web Search and Data Mining}}.
  \bibinfo{pages}{610--618}.
\newblock


\bibitem[Xia et~al\mbox{.}(2008)]%
        {xia2008listwise}
\bibfield{author}{\bibinfo{person}{Fen Xia}, \bibinfo{person}{Tie-Yan Liu},
  \bibinfo{person}{Jue Wang}, \bibinfo{person}{Wensheng Zhang}, {and}
  \bibinfo{person}{Hang Li}.} \bibinfo{year}{2008}\natexlab{}.
\newblock \showarticletitle{Listwise approach to learning to rank: theory and
  algorithm}. In \bibinfo{booktitle}{\emph{Proceedings of the 25th
  international conference on Machine learning}}. \bibinfo{pages}{1192--1199}.
\newblock


\bibitem[Yadav et~al\mbox{.}(2019)]%
        {yadav2019fair}
\bibfield{author}{\bibinfo{person}{Himank Yadav}, \bibinfo{person}{Zhengxiao
  Du}, {and} \bibinfo{person}{Thorsten Joachims}.}
  \bibinfo{year}{2019}\natexlab{}.
\newblock \showarticletitle{Fair learning-to-rank from implicit feedback}.
\newblock \bibinfo{journal}{\emph{arXiv preprint arXiv:1911.08054}}
  (\bibinfo{year}{2019}).
\newblock


\bibitem[Yilmaz et~al\mbox{.}(2020)]%
        {yilmaz2020reliability}
\bibfield{author}{\bibinfo{person}{Emine Yilmaz}, \bibinfo{person}{Nick
  Craswell}, \bibinfo{person}{Bhaskar Mitra}, {and} \bibinfo{person}{Daniel
  Campos}.} \bibinfo{year}{2020}\natexlab{}.
\newblock \showarticletitle{On the Reliability of Test Collections for
  Evaluating Systems of Different Types}.
\newblock \bibinfo{journal}{\emph{arXiv preprint arXiv:2004.13486}}
  (\bibinfo{year}{2020}).
\newblock


\end{thebibliography}

\end{document}